\documentclass[journal]{IEEEtran}
\usepackage[utf8]{inputenc}
\usepackage{cite}
\usepackage{hyperref}
\usepackage{multirow}
\usepackage{algorithmic}
\usepackage{orcidlink}
\usepackage{subfigure}
\usepackage{amsmath} 
\usepackage{amsfonts,amssymb}  
\usepackage{colortbl}
\usepackage{makecell}
\usepackage{xcolor}
\usepackage{color}
\usepackage{threeparttable} 
\usepackage{xcolor}
\usepackage{graphicx}
\graphicspath{ {./figures/} }
\hypersetup{
	colorlinks=true,
	linkcolor=cyan,
	filecolor=blue,      
	urlcolor=black,
	citecolor=green,
}
\begin{document}

\title{Spatial-Temporal-Spectral Mamba with Sparse Deformable Token Sequence for Enhanced MODIS Time Series Classification}

\author{Zack Dewis\textsuperscript{*}, Zhengsen Xu\textsuperscript{*}, Yimin Zhu, Motasem Alkayid, Mabel Heffring, Lincoln Linlin Xu, ~\IEEEmembership{Member,~IEEE} 


\thanks{This work was supported by the Natural
Sciences and Engineering Research Council of Canada (NSERC) under Grant RGPIN-2019-06744.}
\thanks{Zack Dewis, Zhengsen Xu, Yimin Zhu, Mabel Heffring, Lincoln Linlin Xu are all with the Department of Geomatics Engineering, University of Calgary, Canada (email: (zachary.dewis, zhengsen.xu, yimin.zhu, mabel.heffring1, lincoln.xu)@ucalgary.ca) (Corresponding author: Lincoln Linlin Xu; Zack Dewis and Zhengsen Xu worked equally and are the co-first authors.)}.

\thanks{Motasem Alkayid is with the Department of Geomatics Engineering, University of Calgary, Canada, and also with the Department of Geography, Faculty of Arts, The University of Jordan, Amman, Jordan (email: motasem.alkayid@ucalgary.ca)
}}

\markboth{Journal of \LaTeX\ Class Files,~Vol.~13, No.~9, September~2014}
{Shell \MakeLowercase{\textit{et al.}}: }
\maketitle

\begin{abstract}

Although MODIS time series data are critical for supporting dynamic, large-scale land cover land use classification, it is a challenging task to capture the subtle class signature information due to key MODIS difficulties, e.g., high temporal dimensionality, mixed pixels, and spatial-temporal-spectral coupling effect. This paper presents a novel spatial-temporal-spectral Mamba (STSMamba) with deformable token sequence for enhanced MODIS time series classification, with the following key contributions. First, to disentangle temporal-spectral feature coupling, a temporal grouped stem (TGS) module is designed for initial feature learning. Second, to improve Mamba modeling efficiency and accuracy, a sparse, deformable Mamba sequencing (SDMS) approach is designed, which can reduce the potential information redundancy in Mamba sequence and improve the adaptability and learnability of the Mamba sequencing. Third, based on SDMS, to improve feature learning, a novel spatial-temporal-spectral Mamba architecture is designed, leading to three modules, i.e., a sparse deformable spatial Mamba module (SDSpaM), a sparse deformable spectral Mamba module (SDSpeM), and a sparse deformable temporal Mamba module (SDTM) to explicitly learn key information sources in MODIS. The proposed approach is tested on MODIS time series data in comparison with many state-of-the-art approaches, and the results demonstrate that the proposed approach can achieve higher classification accuracy with reduced computational complexity. 

\end{abstract}


\IEEEpeerreviewmaketitle

\section{Introduction}

\begin{figure}[htbp]
    \centering
    \includegraphics[width=0.51\textwidth]{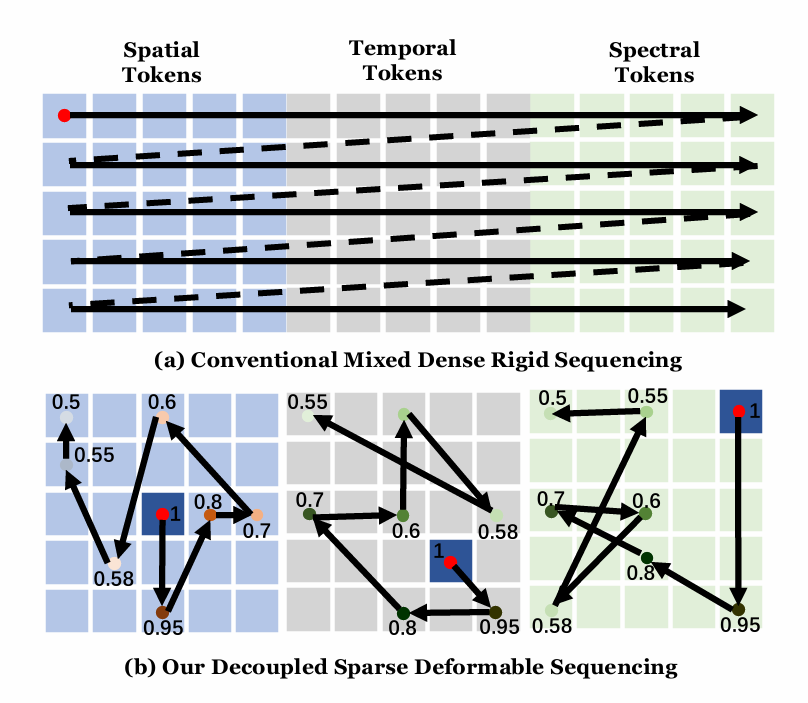}
    \caption{Illustration of (a) conventional \textcolor{red}{\textbf{long, mixed, dense and rigid}} Mamba sequencing and (b) the proposed \textcolor{green}{\textbf{short, decoupled, sparse and deformable}} Mamba sequencing for addressing the spatial-temporal-spectral information coupling in MODIS time series. First, to improve the \textcolor{black}{\textbf{mixed, long}} Mamba sequences in (a), we \textcolor{black}{\textbf{disentangle spatial-temporal-spectral information coupling}} in MODIS with \textcolor{black}{\textbf{three dedicated short Mamba sequences}} in (b). Second, to improve the \textcolor{black}{\textbf{dense, rigid, predefined scanning}} Mamba sequence with all tokens in (a), we design \textcolor{black}{\textbf{sparse, deformable, learnable and adaptive}} Mamba sequence with only most relevant tokens \textcolor{black}{\textbf{to alleviate information redundancy, computational cost, and correlation decay in long Mamba sequences}. 
     }}
    \label{fig:sequence}
\end{figure}
MODIS time series data, due to their high temporal resolution, are critical for supporting dynamic, large-scale land cover and land use (LCLU) classification \cite{FRIEDL2010168}. However, accurate and efficient classification of MODIS time series data is a challenging task due to some key characteristics of MODIS data, i.e., high temporal dimensionality, mixed pixels, and spatial-temporal-spectral coupling effect. First, due to its high temporal resolution, MODIS tends to offer a long time series data with various time steps, leading to a high temporal dimensionality issue that challenges efficient temporal feature learning. Second, MODIS data has coarse spatial resolution, 250m to 1km, leading to various "mixed" pixels where the observed spectral data are a mixture of multiple classes \cite{lobell2004cropland,fisher1997pixel}. These mixed pixels lead to significant spatial heterogeneity and class signature ambiguity \cite{975000}, which poses significant challenges for machine learning (ML) algorithms in learning the appearance of different classes \cite{DENG2013100}. Third, in MODIS, the spatial-temporal-spectral information tends to be coupled together, making it difficult to capture discriminative class signature information of different land cover types \cite{6927890}. Given these difficulties and challenges, advanced ML and deep learning (DL) techniques with enhanced spatial-temporal-spectral feature learning capability are fundamental to improve MODIS time series classification.



Different techniques have been proposed for this purpose.
For example, the support vector machine (SVM) \cite{shao2012comparison, gonccalves2006land, vuolo2012exploiting} and Random Forest (RF) are popular classifiers for MODIS data classification \cite{nguyen2020characterizing, nitze2015temporal, ramo2017developing, yin2014mapping, hao2015feature}, but they struggle with efficient feature engineering approaches to extract meaningful features. Deep learning based feature learning methods, such as CNN \cite{yin2021cascaded, sun2019classification, song2018spatiotemporal}, Transformers \cite{gao2025large, chen2025high, li2021msnet, wang2022spectral} and RNN/LTSM \cite{zhang2022cnn,arslan2019application, ma2022development, mou2018learning} have been widely used for remote sensing and MODIS imagery classification. RNN, especially Long Short-Term Memory (LSTM) models, which are designed to address sequential data, have shown success in remote sensing time series analysis. For example, Ienco \cite{ienco2017land} proposes a LSTM, which improves the classification accuracy for complex and mixed land cover classes. Sun \cite{sun2019land} applies LSTM to Landsat time series classification, achieving accuracies of 97.2\% for five classes and 88.4\% across 132 classes. To overcome the limitation of requiring large labelled datasets, Jing and Chao \cite{jing2020semi} introduces the semi-supervised convolutional LSTM (ConvLSTM), which allows for more robust classification in scenarios with high cloud prevalence or when ground truth is sparse. However, despite their strong temporal modeling capabilities, RNN and LSTM struggle to capture the subtle spatial information and to process long time sequence efficiently.
 
Temporal CNNs apply convolution operations in temporal domain to learn temporal information. For example, Pelletier \cite{pelletier2019temporal} uses TempCNNs on multiple satellite datasets, demonstrating better performance than RNN-based approaches in both accuracy and training speed. Brock and Abdallah \cite{brock2022review} further validate the strength of the Temporal CNN approach, especially for agricultural monitoring, as crops exhibit strong seasonal behavior. Temporal CNNs excel at capturing local temporal features, such as sudden vegetation change or the onset of planting/harvesting phases. However, CNNs rely on local receptive fields that inherently limit their ability to capture long-range dependencies and temporal dynamics in multi-temporal datasets \cite{mou2018learning}. In contrast, Transformer architectures, despite their larger-scale modelling strength, may struggle with the high computational cost, and the inefficiency at addressing the sequential nature of time-series data \cite{basnyat2000use,khan2024transformer}. Recent advancements, such as the Earthformer model, address these issues by incorporating cuboid attention mechanisms, but these adaptations are still evolving and may not fully capture the complexities of temporal relationships in remote sensing data \cite{gao2022earthformer}.


Recently, the Mamba approach has been widely used for remote sensing image classification due to its ability to capture long-range correlation with reduced computational cost. For example, Mamba-based methods have emerged as a promising approach to hyperspectral image (HSI) classification \cite{10604894, 10703171, liu2024hypermamba}, which demonstrate better performances than CNNs and Transformers. These models leverage the state space model (SSM) framework to efficiently capture spatial-spectral dependencies with linear computational complexity \cite{ahmad2024comprehensive}. However, the use of Mamba for MODIS time series classification is insufficiently researched. There are two critical issues that need to be addressed. (1) How to develop dedicated spatial-temporal-spectral Mamba for enhanced feature learning from MODIS time series data; (2) how to improve the Mamba architecture by building a token sequence in a sparse and learnable manner. Addressing these issues is critical for improving MODIS time series classification. 

This paper presents a novel spatial-temporal-spectral Mamba (STSMamba) with deformable token sequence for enhanced MODIS time series classification, with the following contributions. 

\begin{itemize}

\item First, to disentangle temporal-spectral feature coupling, a temporal grouped stem (TGS) module is designed for initial feature learning in the proposed Mamba architecture. This module separates temporal and spectral information and builds the foundation for subsequent modules. 

\item Second, to improve Mamba modeling efficiency and accuracy, a sparse, deformable Mamba sequencing (SDMS) approach is designed, which can reduce the potential information redundancy in Mamba sequence and improve the adaptability and learnability of the Mamba sequencing. As illustrated in Figure. \ref{fig:sequence}, the proposed sparse, deformable, learnable and adaptive Mamba sequencing approach can alleviate information redundancy, computational cost, and correlation decay in long Mamba sequences.

\item Third, based on SDMS, to improve feature learning, a novel spatial-temporal-spectral Mamba architecture is designed, leading to three modules, i.e., a sparse deformable spatial Mamba module (SDSpaM), a sparse deformable spectral Mamba module (SDSpeM), and a sparse deformable temporal Mamba module (SDTM) to explicitly learn key information sources in MODIS. As illustrated in Figure. \ref{fig:sequence}, different with the mixed Mamba sequences, the proposed approach can disentangle spatial-temporal-spectral information coupling in MODIS with three dedicated Mamba modules. 

\end{itemize}

The proposed approach is tested on MODIS MOD13Q1 time series data in comparison with many state-of-the-art classification approaches, i.e., CNN, Transformer and Mamba approaches, and the results demonstrate that the proposed approach can achieve higher classification accuracy with less computational complexity. In addition, extensive ablation studies are conducted to justify the importance and benefits of the key building blocks of the proposed approach. 

The remainder of the paper is organized as follows. Section \ref{section2} talks about the related works. Section \ref{methodology} illustrates the details of the proposed STSMamba approach. Section \ref{results} presents the experimental design and results. Section \ref{conclusion} concludes this study.

\section{Related Works on Sparse and Deformable Models} \label{section2}


Recent advances in machine learning tend to promote sparse models and deformable architectures. Sparse models are inspired by biological systems, such as the principle of selective activation of the brain, where only a small subset of neurons is activated at any given time \cite{sinha2022new}. Sparse model can better address the information redundancy issue to improve model efficiency and reduce computational and memory cost. Moreover, sparse models benefit from improved generalization as the sparsity acts as an implicit regularizer, which prevents overfitting. Sparsity can be achieved in different ways, i.e., local attention, pruning, dynamic sparsity, and learnable sparsity \cite{farina2024sparsity}.

Sparsity is widely used in Transformer models to reduce redundancy in attention matrix \cite{farina2024sparsity}, which can also reduce the computational cost of Transformer models. For example, Child et al., \cite{child2019generating} split the full attention matrix in Transformer into strided attention and fixed attention. This approach allows different attention heads to use their own sparse patterns, but make sure that all positions in the attention matrix are covered. It reduces the attention computation from \(O(N^2)\) to \(O(N \sqrt{N})\). Child et al., Roy et al. employ a similar approach of bound and strided attention, but implement k-means clustering to further increase the efficiency of the attention mechanism \cite{roy2021efficient}. Jaszczur et al., make every key component in Transformer to be sparse, including the feedforward layer, the QKV layer, and the loss layer in natural language processing \cite{jaszczur2021sparse}. An adaptive sparse transformer approach is achieved by making the shape of each attention head learnable to allow greater interpretability and accuracy \cite{correia2019adaptively}. To further reinforce the wide variability of sparse approaches, Pinasthika et al. introduce a sparse transformer block where the final stage of the model extracts critical features through a convolution layer before pixel classification \cite{pinasthika2024sparseswin}. Sparsity is not limited to just transformer models, but is also widely used in other attention-based models. For example, Shirzad, et al replace the transformer architecture with a sparse graph neural network to better capture global and local features through expander graphs edges used as attention patterns \cite{shirzad2023exphormer}. Given the success achieved by sparse Transformer models, it is critical to explore sparsity in Mamba models for improved efficiency and reduced computational cost. 

Meanwhile, deformable models address a fundamental limitation in traditional machine learning, i.e., rigid inductive biases, such as fixed convolutional kernels. Deformable models address this by adapting to input geometry to better captures real-world variability. Additionally, deformable models offer greater parameter efficiency by requiring fewer parameters to model complex geometric transformations. They are also more robust to input distortions, which makes them inherently more invariant and less dependent on extensive data augmentation.  




Deformable approaches are widely used in deep learning models, leading to improved model performance. For example, Zhu et al. find that replacing normal convolution layers with deformable ones and stacking them leads to higher accuracy and efficiency \cite{zhu2019deformable}. Wang et al. design a sparse deformable kernel and stack the blocks to model a more global view, which achieves similar results to ViTs \cite{wang2023internimage}. CNNs are not the only model that benefit from deformability. Similar improvements were found in the transformer attention module, where deformable approaches mitigate the slow convergence and high complexity in  transformers \cite{zhu2020deformable}. Xia et al. use deformable attention module to improve object detection with greater efficiency compared to other vision transformers \cite{xia2022vision}. Jin et al, combine the UNet architecture with deformability and find that the addition of a deformable block enables more detailed features extraction than UNet \cite{jin2019dunet}. 

Given the importance and benefits of sparsity and deformability, it is critical to design sparse and deformable Mamba models to improve modeling efficiency, mitigate the correlation decay issue in long Mamba sequance, and reduce the computational cost and memory consumption in MODIS spatial-temporal-spectral data. 

\section{Methodology} \label{methodology}

\begin{figure*}
    \centering
    \includegraphics[width=1\linewidth]{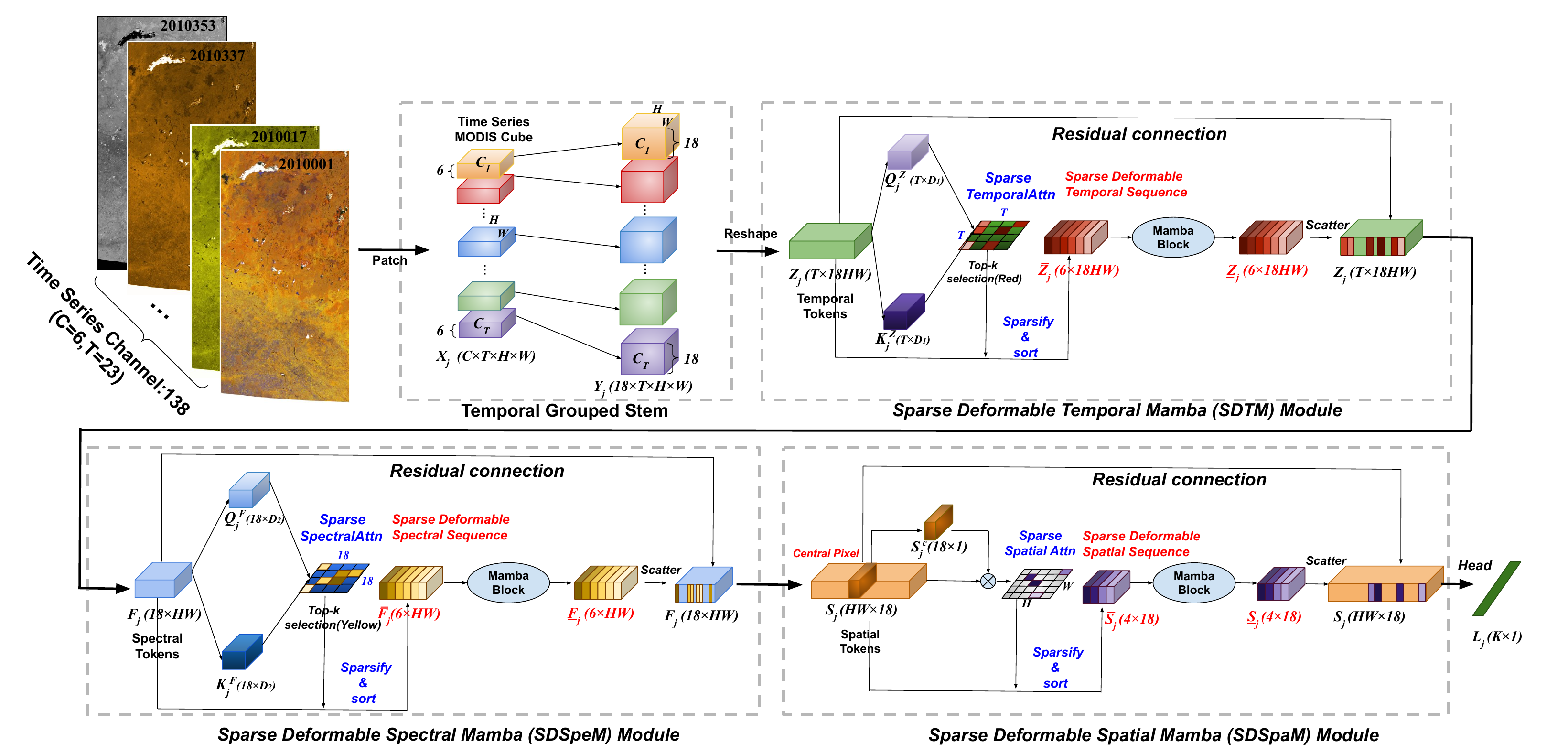}
    \caption{The input MODIS images have 23 time steps, with each step having 6 spectral channels. The proposed STSMamba \textbf{disentangles} this spatial-temporal-spectral information coupling effect via three dedicated modules, i.e., SDTM, SDSpeM and SDSpaM, which are implemented using a novel \textbf{sparse and deformable} Mamba approach. First, SDTM, SDSpeM and SDSpaM have \textbf{sparse} Mamba sequence, because the input sequences to the \textbf{MambaBlock}, i.e.,  $\overline{\boldsymbol{\mathit{Z}}}_j (6 \ tokens)$,  $\overline{\boldsymbol{\mathit{F}}}_j (6 \ tokens)$, and $\overline{\boldsymbol{\mathit{S}}}_j (4 \ tokens)$ have much less number of tokens than  $\boldsymbol{\mathit{Z}}_j (T \ tokens)$,  $\boldsymbol{\mathit{F}}_j (18 \ tokens)$, and $\boldsymbol{\mathit{S}}_j (HW \ tokens)$ respectively. Second, SDTM, SDSpeM and SDSpaM have \textbf{deformable} Mamba sequence, because the order of tokens in $\overline{\boldsymbol{\mathit{Z}}}_j$, $\overline{\boldsymbol{\mathit{F}}}_j$, and $\overline{\boldsymbol{\mathit{S}}}_j$ are deformable and learnable. Three adaptive attention matrices, i.e., $SparseTemporalAttn$, $SparseSpectralAttn$, and $SparseSpatialAttn$ are used to sort the tokens and identify limited number of relevant and informative tokens. Therefore, the proposed STSMamba model has \textbf{disentangled, sparse, and deformable Mamba} sequence that can reduce redundancy, rigidity, and computational cost in classical Mamba models}.
    \label{fig:model architecture}
\end{figure*}

\subsection{Overview}

Figure \ref{fig:model architecture} displays the architecture of the proposed STSMamba model. This model input is \(X_{j} \in \mathbb{R}^{B \times (T \times C) \times H \times W}\), where \(B\), \(T\), \(C\), \(H\), and \(W\) are respectively batch size, the number of time steps, the number of spectral channels, the hight, and the width of the MODIS time series images. For our dataset, \(T\) equals 23, and \(C\) equals 6. 

First, to disentangle temporal-spectral feature coupling, a temporal grouped stem (TGS) module is designed, as illustrated in Section \ref{stem}. 

Second, to achieve disentangled, sparse and deformable Mamba, three modules, i.e., SDTM, SDSpeM, SDSpaM are designed, as illustrated in Sections \ref{SDTM}, \ref{SDSpeM}, and \ref{SDSpaM} respectively.


\subsection{Temporal Group Stem Layer (TGS)}
\label{stem}
Figure \ref{fig:model architecture} indicates that the input MODIS data cube \(X_j \in \mathbb{R}^{C \times T \times H \times W}\), is separated into \(T\) groups, with each group \(C_t \in \mathbb{R}^{C \times H \times W}\) being one time step. 

Instead of using \((T \times C)\) bands simultaneous to feed stem convolution layers, we address each time step $C_t$ individually and use $T$ bands in $C_t$ to feed to the stem convolutions layers, which output 18 features for each time step.   
\begin{align}
    \begin{aligned}
        C_t = \text{GELU}(\text{BN}(\text{2DConv}(C_t)))
    \end{aligned}
\end{align}
where 2Dconv, GELU and BN are $3 \times 3$ 2D convolution kernel, Gaussian Error Linear Units and Batch Normalization, respectively. The global BN ensures consistent distribution for each temporal group. 

\subsection{Sparse Deformable Temporal Mamba Module (SDTM)}
\label{SDTM}

Figure \ref{fig:model architecture} indicates that \(Y_{j} \in \mathbb{R}^{18 \times T \times H \times W} \) is reshaped into \(Z_{j} \in \mathbb{R}^{T \times 18HW}\), where \(T\) is the number of temporal tokens, with each token being a \(18HW \times 1\) spatial-spectral vector. 

Instead of using $Z_j$ with $T$ tokens to feed \textit{MambaBlock}, to achieve sparse and deformable Mamba sequence, we generate $\overline{\boldsymbol{\mathit{Z}}}_j (6 \ tokens)$ with only six re-ordered tokens to feed \textit{MambaBlock}. 

How to identify these six tokens? We use a \textit{SparseTemporalAttn} approach. We first calculate \textit{TemAM} and then sparsify it to achieve \textit{SparseTemporalAttn}. 

The initial Temporal attention matrix (\textit{TemAM} \(\in \mathbb{R}^{T \times T}\)) can be expressed as follows:
\begin{align}
    \begin{aligned}
        TemAM = Attention(\mathcal{Q}_j, \mathcal{K}_j) = \sigma(\frac{\mathcal{Q}_j \mathcal{K}^T_j}{\sqrt{D}})
    \end{aligned}
    \label{EQ_sparseattn}
\end{align}
where \(\mathcal{Q}_j = Z_j \mathcal{W}^{\mathcal{Q}_j} \in \mathbb{R}^{T\times D} \), \(\mathcal{K}_j = Z_j \mathcal{W}^{\mathcal{K}_j} \in \mathbb{R}^{T\times D}\) are queries, keys of temporal tokens, with \(D\) being the hidden dimension. \(\mathcal{W}^{\mathcal{Q}_j}\) and \(\mathcal{W}^{\mathcal{K}_j}\)
are the projection weights of \(\mathcal{Q}_j\), \(\mathcal{K}_j\), and \(\sigma\) is Softmax function. 


Based on \textit{TemAM}, we achieve \textit{SparseTemporalAttn} by:
\begin{align}
    \begin{aligned}
        MeanVec &= [\mu_1, ..., \mu_T], with \ \mu_i = \frac{1}{T}\underset{n=1}{\overset{T}{\Sigma}} TemAM_{mn}\\
        SortedMeans&=\text{sort}(MeanVec,\text{descending=True})\\
        index&=SortedMeans(0: \lfloor \lambda \times T \rfloor)\\
        \bar{Z}_j&=Z_j(index)
    \end{aligned}
    \label{EQ_Tem}
\end{align}
where, \(\lambda\) is the sparse ratio, and \(\lambda \times T\) gives the number of tokens in the sparse Mamba sequence. Equation \ref{EQ_Tem} not only gives sparse Mamba sequence, but also provides deformable and learnable \(\bar{Z}_j\), because \textit{TemAM} is learnable and \textit{SortedMeans} is deformable. 

We use sparse and deformable \(\bar{Z}_j\) as input to \textit{MambaBlock}. The output of \textit{MambaBlock}, denoted by $\underline{\boldsymbol{\mathit{Z}}}_j$, is scattered into the temporal dimensions of $Z_j$, which serves as a residual skip connection. 

\subsection{Sparse Deformable Spectral Mamba Module (SDSpeM)}
\label{SDSpeM}


Figure \ref{fig:model architecture} indicates that \(Z_j \in \mathbb{R}^{T \times 18HW}\) is reshaped into \(F_j \in \mathbb{R}^{T \times 18 \times HW}\), leading to a total of 18 spectral tokens, with each token being a \(HW \times 1\) spatial vector. Here, the temporal dimension $T$ is treated as the batch dimension, and thereby there are a total of $T$ samples, with each sample owning 18 tokens. 

Similar to Section \ref{SDTM}, to achieve sparse and deformable Mamba, instead of using $F_j$ with $18$ tokens to feed \textit{MambaBlock}, we generate $\overline{\boldsymbol{\mathit{F}}}_j (6 \ tokens)$ with only six re-ordered tokens to feed \textit{MambaBlock}. 

How to identify these six tokens in $\overline{\boldsymbol{\mathit{F}}}_j (6 \ tokens)$? We use a \textit{SparseSpectralAttn} approach. We first calculate \textit{SpecAM} and then sparsify it to achieve \textit{SparseSpectralAttn}. 

To achieve $\overline{\boldsymbol{\mathit{F}}}_j (6 \ tokens)$, the Spectral attention matrix (\textit{SpecAM} \(\in \mathbb{R}^{18 \times 18}\)) is first calculated in the same way as in Equation \ref{EQ_sparseattn}, based on which,  $\overline{\boldsymbol{\mathit{F}}}_j (6 \ tokens)$  can be obtained in a similar manner as in Equation \ref{EQ_Tem}.

We use sparse and deformable $\overline{\boldsymbol{\mathit{F}}}_j (6 \ tokens)$ as input to \textit{MambaBlock}. The output of \textit{MambaBlock}, denoted by $\underline{\boldsymbol{\mathit{F}}}_j$, is scattered into the temporal dimensions of $F_j$, which serves as a residual skip connection. 

\subsection{Sparse Deformable Spatial Mamba Module (SDSpaM)}
\label{SDSpaM}

Figure \ref{fig:model architecture} indicates that \(F_j \in \mathbb{R}^{18 \times HW}\) is reshaped into \(S_j \in \mathbb{R}^{HW \times 18}\), leading to a total of HW spatial tokens, with each token being a \(18 \times 1\) spectral vector.

Similar to Section \ref{SDTM} and \ref{SDSpeM}, to achieve sparse and deformable Mamba, instead of using $S_j$ with $HW$ tokens to feed \textit{MambaBlock}, we generate $\overline{\boldsymbol{\mathit{S}}}_j (4 \ tokens)$ with only four re-ordered tokens to feed \textit{MambaBlock}. 

How to identify these four tokens in $\overline{\boldsymbol{\mathit{S}}}_j (4 \ tokens)$? We use a \textit{SparseSpatialAttn} approach. We first calculate \textit{SpatialAttn} and then sparsify it to achieve \textit{SparseSpatialAttn}. 

The spatial attention matrix (\textit{SpatialAttn} \(\in \mathbb{R}^{H \times W}\)) is first calculated by




\begin{align}
    SpatialAttn_i &= \arccos \left( \frac{S_{i}^{T} S^c}{\| S_{i}\| \| S^c\|} \right) 
\end{align}

\noindent where $SpatialAttn_i$ is the $i$th element of \textit{SpatialAttn}, \(S^{c}\) is the central pixel in the feature map, \(S^{i}\) is the $i$th neighbour pixel in the feature map, and \textit{arccos} measures the similarity between \(S^{c}\) and \(S^{i}\). 

Based on \textit{SpatialAttn}, to achieve \textit{SparseSpatialAttn}, we sort and select top elements in \textit{SpatialAttn}: 
\begin{align}
    \begin{aligned}
        \textit{SparseSpatialAttn} =TopK(\textit{sort}(\textit{SpatialAttn})) \\
    \end{aligned}
    \label{EQ_Spe}
\end{align}

\noindent where \textit{TopK} identifies the top $K=\lambda \times HW$ elements in sorted \textit{SpatialAttn}, and sets the rest of the elements to be zero. We use a sparsity ratio of \(\lambda = 0.3\). 

To achieve $\overline{\boldsymbol{\mathit{S}}}_j (4 \ tokens)$, we follow

\begin{align}
    \begin{aligned}
        index&=sort(NonZeros(SparseSpatialAttn))\\
        \bar{S}_j&=S_j(index)
    \end{aligned}
    \label{EQ_Tem}
\end{align}

We use sparse and deformable \(\bar{S}_j\) as input to \textit{MambaBlock}. The output of \textit{MambaBlock}, denoted by $\underline{\boldsymbol{\mathit{S}}}_j$, is scattered into the temporal dimensions of $S_j$, which serves as a residual skip connection. 



\section{Results and Analysis} \label{results}

\subsection{Datasets}
To test the proposed model, 250m MODIS time series product of year 2010, i.e., MOD13Q1, which covers the Canadian province of Saskatchewan, is adopted. The MOD13Q1 product has a 16-day revisit cycle, leading to 23 time steps in a year. The 30m land cover and land use maps published by Natural Resources Canada (NRCan) is used as ground-truth \cite{maps}. This map is resampled to the 250m resolution to be consistent with the MODIS data.

To test the spatial generalization capability of the proposed model, another MOD13Q1 dataset covering the adjacent province, i.e., Alberta, is adopted to be predicted by the model trained on the Saskatchewan dataset. 

Overall, the two datasets share many similarities but have several key differences. Both provinces feature dominant land cover types such as forest, croplands, grasslands and wetlands, which are common in the prairies and boreal regions of Canada. Both provinces have sparsely populated regions with significant agricultural and natural vegetation coverage, making them a challenge for classification. However, the Saskatchewan dataset has a higher proportion of cultivated land compared to Alberta, where forest and grasslands are more dominant. Alberta's land cover is influenced by the Rocky Mountains (which contains alpine vegetation and snow cover), whereas Saskatchewan is predominantly flat with wetland systems playing a larger role. Finally, Alberta has more pronounced human-altered landscapes due to the oil sands and urban expansion, whereas Saskatchewan is much more agricultural driven. These differences can help test generalization capabilities of the model.

\begin{figure*}
    \centering
    \includegraphics[width=1\textwidth]{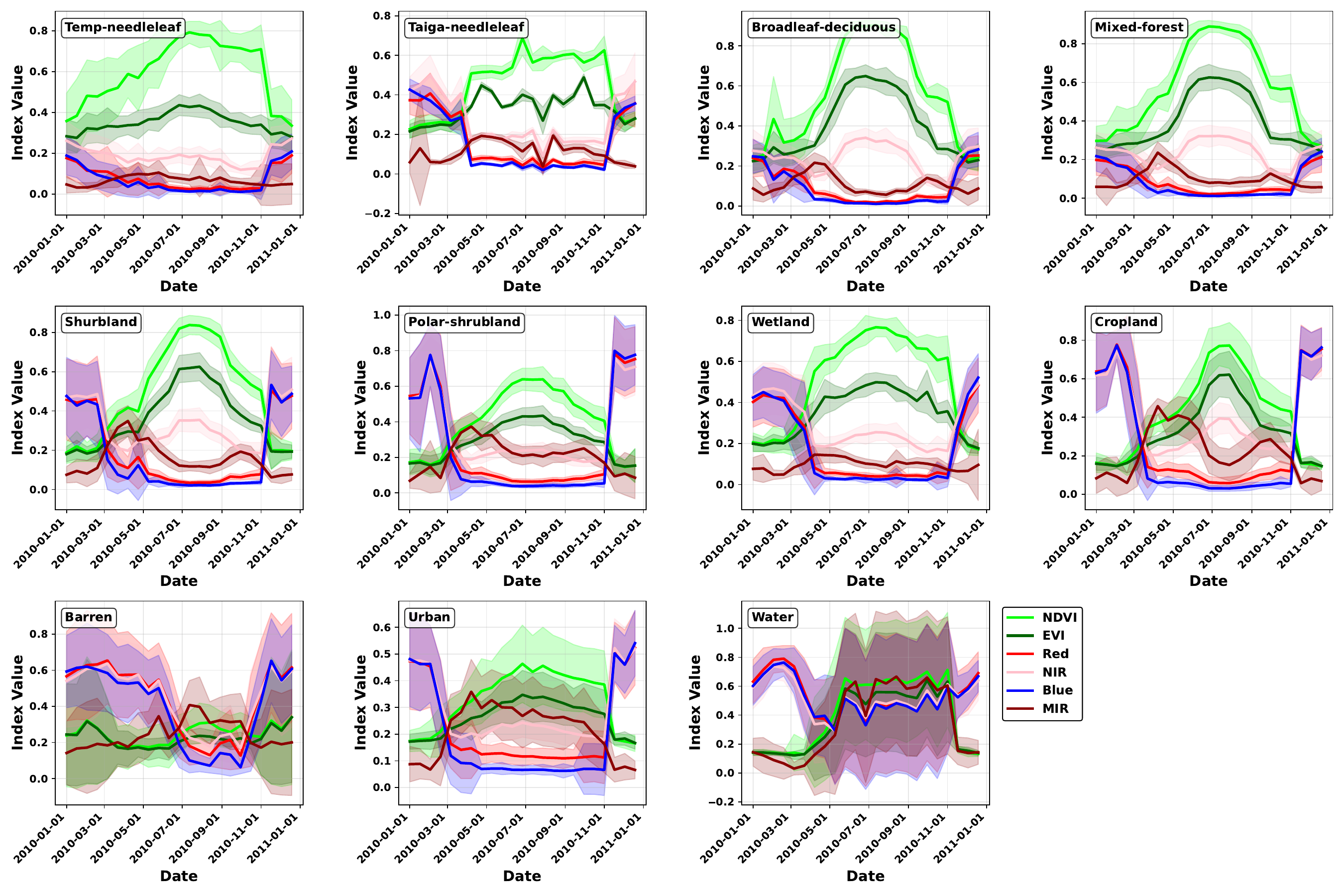}
    \caption{The mean value spectral-temporal curves of different classes in the train Saskatchewan dataset (in total 23 time steps and 6 different spectral bands). It shows the spectral curves throughout the year for the classes in the Saskatchewan dataset. NDVI often sees a peak in the Summer season, as that is when vegetation coverage is highest; this remains true for EVI. In contrast, the red, blue, and NIR bands see a peak in the winter months, due to the high reflectance of ice and snow, which dominates the Canadian winter. These differences in spectral bands in terms of seasonality patterns indicate \textbf{the importance of decoupling the spectral and temporal dimensions} to better highlight the differences. In addition, strong similarities in the spectral-temporal curves occur for multiple classes. For example, Cropland, Shrubland and Polar-shrubland have similar curves. Therefore, \textbf{the model needs to have strong subtle feature extraction capabilities} to be able to differentiate between subtle spectral differences that occur throughout the year to achieve high accuracy.}
    \label{fig:Spectral Curve}
\end{figure*}

Figure \ref{fig:Spectral Curve} shows the spectral curves throughout the year for the classes in the Saskatchewan dataset. NDVI often sees a peak in the Summer season, as that is when vegetation coverage is highest; this remains true for EVI. In contrast, the red, blue, and NIR bands see a peak in the winter months, due to the high reflectance of ice and snow, which dominates the Canadian winter. These differences in spectral bands in terms of seasonality patterns indicate the importance of decoupling the spectral and temporal dimensions to better highlight the differences. In addition, strong similarities in the spectral-temporal curves occur for multiple classes. For example, Cropland, Shrubland and Polar-shrubland have similar curves. Therefore, the model needs to have strong subtle feature extraction capabilities to be able to differentiate between subtle spectral differences that occur throughout the year to achieve high accuracy. 

\begin{table}[]
\caption{Number of Samples in  Saskatchewan \& Alberta MOD13Q1 dataset}
\resizebox{0.49\textwidth}{!}{
\centering
\begin{tabular}{c|cc|cccc|c}
\hline
\multicolumn{1}{c|}{\multirow{2}{*}{Color}} & \multicolumn{1}{c}{\multirow{2}{*}{Class Number}} & \multicolumn{1}{c|}{\multirow{2}{*}{Class Name}} & \multicolumn{4}{c|}{Saskatchewan}    & Alberta       \\
\multicolumn{1}{c|}{}                       & \multicolumn{1}{c}{}                              & \multicolumn{1}{c|}{}                            & Train & Val  & Test  & Final Mapping & Final Mapping \\ \hline
 \cellcolor[RGB]{1, 62, 2}       & 1                                                 & \multicolumn{1}{c|}{Temp-needleleaf} & 100   & 100  & 1000  & 2491052        & 3013683        \\
 \cellcolor[RGB]{149, 156, 112}            & 2       &    \multicolumn{1}{c|}{Taiga-needleleaf}       & 100   & 100  & 1000  & 119247          & 8138            \\
\cellcolor[RGB]{20, 139, 61} & 3                                                 &    \multicolumn{1}{c|}{Temp-needleleaf}                                              & 100   & 100  & 1000  & 531086        & 73627          \\
  \cellcolor[RGB]{93, 117, 43}      & 4                                                  &      \multicolumn{1}{c|}{Mixed forest}                                            & 100   & 100  & 1000  & 185260         & 1989029        \\
 \cellcolor[RGB]{179, 137, 51}    & 5                                                 &            \multicolumn{1}{c|}{Shrubland}                                       & 100   & 100  & 1000  & 317655          & 461360         \\
\cellcolor[RGB]{226, 206, 136}    & 6                                                 &            \multicolumn{1}{c|}{Polar-shrubland}                                       & 100   & 100  & 1000  & 1214870       & 1107038        \\
\cellcolor[RGB]{108, 163, 138}    & 7                                                  &              \multicolumn{1}{c|}{Wetland}                                    & 100   & 100  & 1000  & 707054         & 1052896        \\
\cellcolor[RGB]{231, 174, 103}    & 8                                                  &              \multicolumn{1}{c|}{Cropland}                                    & 100   & 100  & 1000  & 3422675       & 2180954        \\
 \cellcolor[RGB]{166, 171, 174}    & 9                                                 &         \multicolumn{1}{c|}{Bare}                                          & 100   & 100  & 1000  & 233334         & 222537         \\
\cellcolor[RGB]{221, 32, 38}    & 10                                                &          \multicolumn{1}{c|}{Urban}                                        & 100   & 100  & 1000  & 40997         & 88583         \\
 \cellcolor[RGB]{76, 112, 164}   & 11                                                 &                 \multicolumn{1}{c|}{Water}                                 & 100   & 100  & 1000  & 1174863        & 396098        \\ \hline
    & \multicolumn{2}{c|}{Total}                                               & 1100  & 1100 & 11000 & 10397096       & 10593943     \\ \hline   
\end{tabular}
}
\label{train_num}
\end{table}

\subsection{Experimental Settings}
Table \ref{train_num} shows the number of samples. The proposed model is trained on the Saskatchewan dataset, using 100 training samples in each class. Each sample is a \(13 \times 13\) image patch of \(23 \times 6\) temporal-spectral channels. The number of validation and test samples are 100 and 1000 respectively, for each class. To make sure the samples are homogeneous, and to reduce the presence of mixed pixels, we use a 5x5 filter to identify and use pixels whose class labels are the same as their neighbors in the filter. 

For visual evaluation, we generate the final Saskatchewan maps by using the trained model to predict all pixels in the Saskatchewan dataset. 

To test the spatial generalization capability of classifiers, we use an adjacent province, i.e., Alberta, to obtain test accuracies and final Alberta maps. To generate test accuracies, we identify about 1000 samples for each class and use them to test the classifiers. To generate final Alberta maps, the classifiers are used to predict all pixels in the Alberta dataset.  

A total of nine state-of-the-art deep learning models are compared with the proposed method. These models cover the main deep learning categories, i.e., CNN, RNN, LSTM, Transformer, and Mamba approaches. 

All training and testing was performed on a NVIDIA RTX A6000 Ada Generation with 48GB of VRAM using the PyTorch library. The batch size and epoch number are respectively 1024 and 100.

\subsection{Comparison Results}

Tables II indicate that the proposed STSMamba model greatly outperforms the other state-of-the-art Mamba, Transformer, and RNN-based methods across all metrics on the Saskatchewan dataset. STSMamba achieves a 3–8\% increase in OA, AA, and Kappa coefficient compared to a recent Mamba model (i.e., MambaHSI). Comparing with the famous Transformer methods (i.e., SwinT), STSMamba increase OA, AA and Kappa by 3.8\%, 6.2\%, and 4.0\% respectively. Similarly, it outperforms the ViT model. STSMamba surpasses the top RNN-based approach (LSTM) by 3.69 (OA), 3.69 (AA), and 4.06 (Kappa) percentage points. These quantitative results underscore STSMamba’s efficiency in capturing subtle land cover signatures, due to its improved feature learning capabilities. 

\begin{figure*}
    \centering
    \includegraphics[width=1\linewidth]{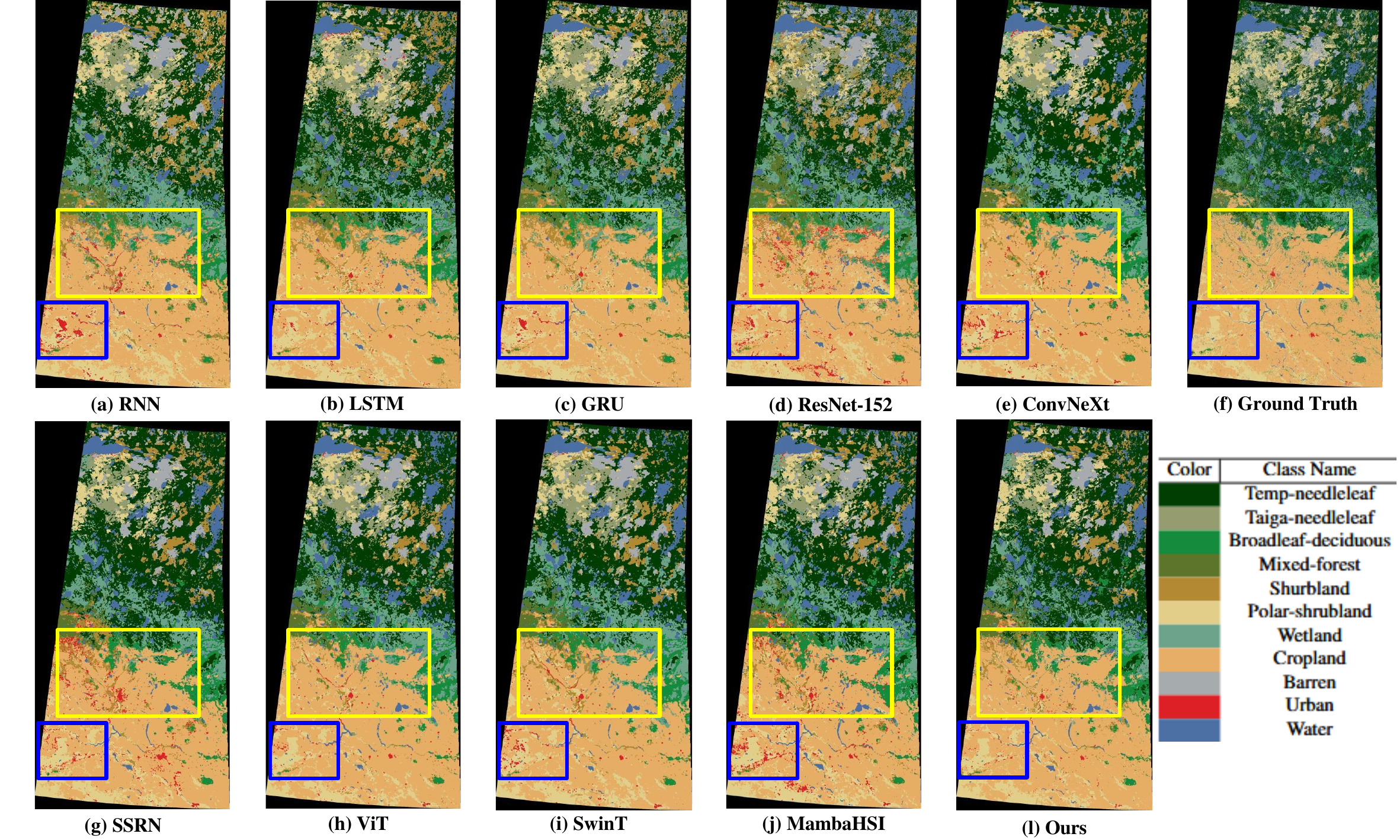}
    \caption{Classification map (250m resolution) of the Saskatchewan dataset. The yellow and blue box show the differences between the methods.}
    \label{Sask map}
\end{figure*}

Visual analysis further validates these findings, as illustrated in Figure \ref{Sask map}. Although the Saskatchewan dataset presents challenging and heterogeneous landscapes, STSMamba consistently outperforms other state-of-the-art models by better classifying subtle classes with sharper segmentation boundaries. For example, the highlighted boxes in Figure \ref{Sask map} indicate that STSMamba can better identify urban regions from non-urban than the other methods.

\begin{table*}[]
\centering
\Large
\caption{Classification results on Filtered ground truth on the Saskatchewan dataset. The best results are in bold.}
\resizebox{\textwidth}{!}{
\begin{tabular}{ccc|ccccccccc|c}
\hline
\multicolumn{1}{c|}{Color} & \multicolumn{1}{c|}{Class Name}          & Class Number & RNN   & LSTM  & GRU   & ResNet-152 & ConvNeXt & SSRN  & ViT   & SwinT & MambaHSI &  Ours  \\ \hline
\cellcolor[RGB]{1, 62, 2}      & \multicolumn{1}{c|}{Temp-needleleaf}     & 1            & 89.67 & 90.85 & 89.57 & 84.35      & 93.29    & 96.26 & 90.66 & 90.99           & 92.63          & \cellcolor[RGB]{251, 228, 213}\textbf{97.41} \\
\cellcolor[RGB]{149, 156, 112}      & \multicolumn{1}{c|}{Taiga-needleleaf}    & 2            & 98.1  & 96.63 & 98.41 & 97.91      & 98.9     & \cellcolor[RGB]{251, 228, 213}\textbf{99.45} & 98.07 & 97.57           & 98.96          & 99.39 \\
\cellcolor[RGB]{20, 139, 61}      & \multicolumn{1}{c|}{Broadleaf-deciduous} & 3            & 91.1  & 89.97 & 90.17 & 75.41      & 94.31    & 95.58 & 93.23 & 89.09           & 93.93          & \cellcolor[RGB]{251, 228, 213}\textbf{97.27} \\
\cellcolor[RGB]{93, 117, 43}     & \multicolumn{1}{c|}{Mixed-forest}        & 4            & 92.76 & 96.45 & 95.42 & 93.43      & 92.89    & 95.54 & 92.34 & 93.63           & 87.62    & \cellcolor[RGB]{251, 228, 213}\textbf{96.09}        \\
\cellcolor[RGB]{179, 137, 51}       & \multicolumn{1}{c|}{Shurbland}           & 5            & 85.34 & 87.9  & 86.13 & 85.81      & 88.45    & \cellcolor[RGB]{251, 228, 213}\textbf{93.35} & 85.62 & 85.5            & 89.01         & 91.91 \\
\cellcolor[RGB]{226, 206, 136}      & \multicolumn{1}{c|}{Polar-shrubland}     & 6            & 73.63 & 87.63 & 82.65 & 74.83      & 88.35    & 92.89 & 87.2  & 84.36           & 90.64          & \cellcolor[RGB]{251, 228, 213}\textbf{93.73} \\
\cellcolor[RGB]{108, 163, 138}       & \multicolumn{1}{c|}{Wetland}             & 7            & 90.18 & 94.25 & 94.61 & 92.99      & 96.87    & 97.54 & 94.14 & 93.31           & 94.83          & \cellcolor[RGB]{251, 228, 213}\textbf{98.12} \\
\cellcolor[RGB]{231, 174, 103}      & \multicolumn{1}{c|}{Cropland}            & 8            & 88.94 & 92.85 & 92.53 & 85.03      & 96.91    & 96.27 & 97.22 & 95.22           & 92.3           & \cellcolor[RGB]{251, 228, 213}\textbf{97.96} \\
\cellcolor[RGB]{166, 171, 174}      & \multicolumn{1}{c|}{Barren}               & 9            & 93.42 & 94.03 & 93.44 & 91.61      & 94.57    & \cellcolor[RGB]{251, 228, 213}\textbf{95.97} & 94.94 & 95.17           & 94.83         & 94.82 \\
\cellcolor[RGB]{221, 32, 38}     & \multicolumn{1}{c|}{Urban}               & 10           & 94.64 & 95.25 & 95.95 & 97.03      & 98.44    & \cellcolor[RGB]{251, 228, 213}\textbf{99.66} & 99.05 & 98.28           & 98.71          & 98.19 \\
\cellcolor[RGB]{76, 112, 164}     & \multicolumn{1}{c|}{Water}               & 11           & 90.6  & 96.66 & 96.55 & 98.62      & 98.01    & 99.1  & 97.83 & 97.5            & 99.21          & \cellcolor[RGB]{251, 228, 213}\textbf{99.43} \\ \hline
\multicolumn{3}{c|}{OA(\%)}                                                          & 88.01 & 92.51 & 91.68 & 85.61      & 95.64    & 96.28 & 95.26 & 93.72           & 93.04          & \cellcolor[RGB]{251, 228, 213}\textbf{97.59} \\
\multicolumn{3}{c|}{AA(\%)}                                                          & 81.08 & 87.95 & 86.63 & 77.68      & 92.84    & \cellcolor[RGB]{251, 228, 213}\textbf{96.51} & 92.19 & 89.77           & 93.88          & 96.01 \\
\multicolumn{3}{c|}{Kappa(\%)}                                                       & 89.85 & 92.95 & 92.31 & 88.82      & 94.63    & 93.91 & 93.67 & 92.78           & 88.86          & \cellcolor[RGB]{251, 228, 213}\textbf{96.76} \\ \hline
\end{tabular}
}
\end{table*}

\subsection{Spatial Generalization Capability Evaluation}
The Alberta dataset is used to test the generalization capabilities of models trained on the Saskatchewan dataset. 

Consistent with Table II, Tables III shows that the proposed STSMamba model greatly outperforms the other state-of-the-art methods on the Alberta dataset. STSMamba outperforms MambaHSI by about 7.2, 12.4, and 4.4 percentage points in terms of OA, AA, and Kappa coefficient respectively. Comparing with SwinT, STSMamba increases OA, AA and Kappa by 7.0, 8.4, and 21.1 percentage points respectively. STSMamba also outperforms the rest of the models, demonstrating the stronger generalization capability when transferring from the Saskatchewan dataset to the Alberta dataset. We also notice that all methods in Table III tend to achieve lower accuracies than Table II, which is reasonable considering the discrepancies between the two dataset. 

\begin{figure*}[h]
    \centering
    \includegraphics[width=1\linewidth]{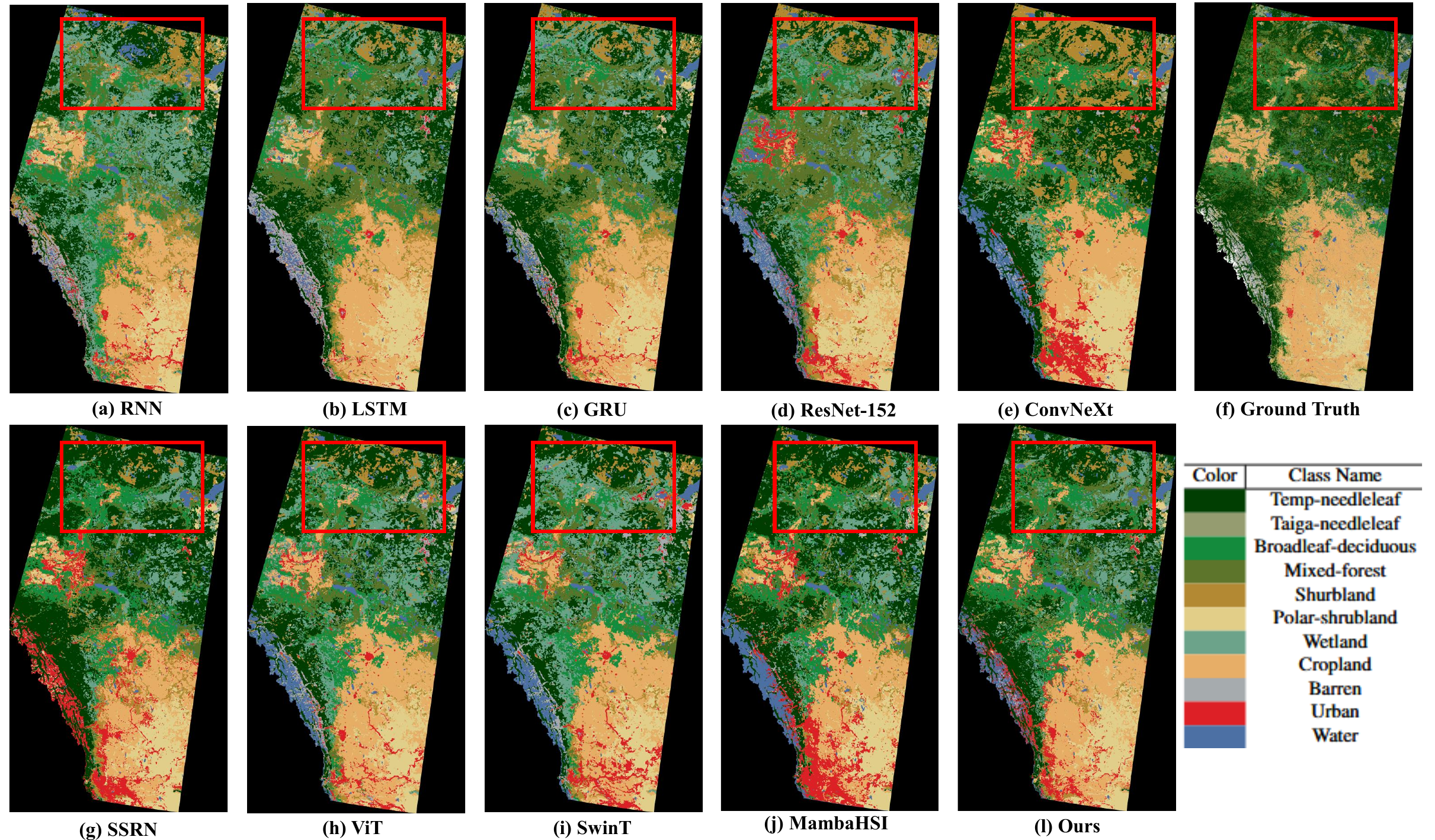}
    \caption{Classification map of Alberta. The red box highlights the predicted differences in the northern part of the province}
    \label{alberta map}
\end{figure*}

Figure \ref{alberta map} shows the Alberta classification maps achieved by different mothods. It indicates that the proposed model generally achieve better map that is more consistent with the ground-truth than the other methods, especially in the upper part of the image, where class signatures are more subtle due to the mixed pixels caused by the presence of various mixed or transitional classes (shrublands, wetlands). 



\subsection{Sensitivity to Sparsity Ratio}

Table \ref{ablation} reveals the influence of sparsity ratios on the performance of the proposed sparse deformable Mamba model.  It indicates that a temporal sparsity ratio of 0.3 achieves optimal performance, likely because this ratio allows the preservation of essential phenological patterns without excessive redundancy. In contrast, spectral sparsity performs slightly better at 0.8, indicating that spectral information richness is critical and aggressive spectral compression could reduce sufficient discriminative power for MODIS data. 


\begin{table}[]
\LARGE 
\caption{Ablation study on Sparsity ratio. The best results are in red font.}
\resizebox{0.49\textwidth}{!}{
\begin{tabular}{c|c|ccc}
\hline
\multicolumn{1}{l|}{}                   &     & \multicolumn{3}{c}{Spectral sparse ratio}              \\ \hline &     & 0.8 & 0.5 & 0.3   \\
& 0.8 & 96.61\textbackslash{}96.61\textbackslash{}96.27                        & 96.46\textbackslash{}96.46/96.11                & 96.70\textbackslash{}96.70\textbackslash{}96.37                        \\
 & 0.5 & 96.62\textbackslash{}96.62\textbackslash{}96.28                        & 96.53\textbackslash{}96.53\textbackslash{}96.18 & 96.73\textbackslash{}96.73\textbackslash{}96.40                        \\
\multirow{-4}{*}{Temporal sparse ratio} & 0.3 & {\color[HTML]{FF0000} \textbf{96.77}\textbackslash{}\textbf{96.77}\textbackslash{}\textbf{96.45}} & 96.80\textbackslash{}96.80\textbackslash{}96.48 & {\color[HTML]{FF0000} \textbf{96.56}\textbackslash{}\textbf{96.56}\textbackslash{}\textbf{96.22}} \\ \hline
\end{tabular}
}
\label{ablation}
\end{table}

\begin{table*}[h]
\centering
\Large
\caption{Spatial Generalization results on Filtered ground truth on the Alberta dataset. The best results are in bold and color shadow.}
\resizebox{\textwidth}{!}{
\begin{tabular}{ccc|ccccccccc|c}
\hline
\multicolumn{1}{c|}{Color} & \multicolumn{1}{c|}{Class Name}          & Class Number & RNN   & LSTM  & GRU   & ResNet-152 & ConvNeXt & SSRN  & ViT   & SwinT & MambaHSI &  Ours  \\ \hline
\cellcolor[RGB]{1, 62, 2}       & \multicolumn{1}{c|}{Temp-needleleaf}     & 1            & 74.42 & 85.46 & 89.76 & 76.5       & 90.32    & 96.07 & 84.2  & 84.04 & 86.09    & \cellcolor[RGB]{251, 228, 213}\textbf{92.28}       \\
\cellcolor[RGB]{149, 156, 112}       & \multicolumn{1}{c|}{Taiga-needleleaf}    & 2            & 90.00    & 90.00    & 90.00    & 85.00         & 90.00       & 90.00    & 37.5  & 7.50   & \cellcolor[RGB]{251, 228, 213}\textbf{92.50}          & 90.00    \\
\cellcolor[RGB]{20, 139, 61}      & \multicolumn{1}{c|}{Broadleaf-deciduous} & 3            & 43.68 & 53.44 & 40.74 & 35.71      & 46.16    & \cellcolor[RGB]{251, 228, 213}\textbf{92.90}  & 30.25 & 21.04 & 60.14         & 66.76 \\
\cellcolor[RGB]{93, 117, 43}      & \multicolumn{1}{c|}{Mixed-forest}        & 4            & 39.13 & 72.88 & 72.31 & \cellcolor[RGB]{251, 228, 213}\textbf{75.37}      & 47.51    & 35.71 & 53.14 & 55.16 & 40.27         & 30.75 \\
\cellcolor[RGB]{179, 137, 51}    & \multicolumn{1}{c|}{Shurbland}           & 5            & 27.06 & 51.9  & 61.96 & 60.25      & 55.69    & \cellcolor[RGB]{251, 228, 213}\textbf{71.08} & 46.8  & 41.37 & 35.68         & 67.20 \\
\cellcolor[RGB]{226, 206, 136}     & \multicolumn{1}{c|}{Polar-shrubland}     & 6            & 72.61 & 85.54 & 84.27 & 78.88      & 75.69    & \cellcolor[RGB]{251, 228, 213}\textbf{92.77} & 80.89 & 69.38 & 85.88          & 90.97 \\
\cellcolor[RGB]{108, 163, 138}     & \multicolumn{1}{c|}{Wetland}             & 7            & 32.26 & 49.10  & 49.10  & 44.26      & 44.10     & 54.18 & 41.74 & 43.32 & 43.20          & \cellcolor[RGB]{251, 228, 213}\textbf{63.24} \\
\cellcolor[RGB]{231, 174, 103}      & \multicolumn{1}{c|}{Cropland}            & 8            & 74.57 & 82.72 & 81.18 & 61.99      & 86.88    & 74.91 & 90.85 & 84.98 & 79.13    & \cellcolor[RGB]{251, 228, 213}\textbf{91.46}      \\
\cellcolor[RGB]{166, 171, 174}      & \multicolumn{1}{c|}{Barren}               & 9            & \cellcolor[RGB]{251, 228, 213}\textbf{63.16} & 34.16 & 17.05 & 45.36      & 10.78    & 4.77  & 26.07 & 18.59 & 3.32          & 23.51 \\
\cellcolor[RGB]{221, 32, 38}       & \multicolumn{1}{c|}{Urban}               & 10           & 70.06 & 75.04 & 70.47 & 93.51      & 85.15    & \cellcolor[RGB]{251, 228, 213}\textbf{97.13} & 83.02 & 67.67 & 94.14         & 96.01 \\
\cellcolor[RGB]{76, 112, 164}     & \multicolumn{1}{c|}{Water}               & 11           & 78.05 & 91.05 & 92.32 & 89.33      & 87.77    & 94.43 & 82.50  & 82.68 & 95.73          & \cellcolor[RGB]{251, 228, 213}\textbf{95.99} \\ \hline
\multicolumn{3}{c|}{OA(\%)}                                                          & 68.03 & 80.40  & 80.45 & 70.04      & 78.12    & 77.55 & 79.23 & 75.36 & 75.09    & \cellcolor[RGB]{251, 228, 213}\textbf{82.36}       \\
\multicolumn{3}{c|}{AA(\%)}                                                          & 60.68 & 75.36 & 75.39 & 63.48      & 72.36    & 73.27 & 73.67 & 69.10  & 65.10     & \cellcolor[RGB]{251, 228, 213}\textbf{77.55}      \\
\multicolumn{3}{c|}{Kappa(\%)}                                                       & 60.45 & 70.12 & 68.10  & 67.83      & 64.46    & 72.06 & 59.72 & 52.34 & 69.03         & \cellcolor[RGB]{251, 228, 213}\textbf{73.47} \\ \hline
\end{tabular}
}
\end{table*}


\section{Conclusion} \label{conclusion}
This paper has presented a novel spatial-temporal-spectral Mamba (STSMamba) with sparse deformable token sequence for enhanced MODIS time series classification. First, a temporal grouped stem (TGS) module was designed to disentangle temporal-spectral feature coupling. Second, a sparse, deformable Mamba sequencing (SDMS) approach was designed to improve Mamba modeling efficiency and accuracy. Third, a novel spatial-temporal-spectral Mamba architecture was designed to improve feature learning. The proposed approach was tested on MODIS time series data in comparison with many state-of-the-art approaches, and the results demonstrated that the proposed approach can achieve higher classification accuracy with reduced computational complexity. Future research directions include using domain shift to further improve the model's generalization capability and using a higher-resolution dataset, e.g., Sentinel-2 to improve small class features.

\bibliographystyle{IEEEtran}
\bibliography{IEEEabrv,references}

\end{document}